%% file: main.tex
\documentclass[final,5p,times,twocolumn]{elsarticle}
\usepackage[utf8]{inputenc}
\usepackage{amsmath}
\usepackage{graphicx}
\usepackage{physics}
\usepackage{siunitx}
\usepackage{enumerate}
\usepackage{tikz}
\usepackage{xcolor}
\usepackage{tikz-3dplot}
\usepackage{hyperref}
\hypersetup{colorlinks = true,linkcolor = blue,anchorcolor =red,citecolor = blue,filecolor = red,urlcolor = red,
            pdfauthor=author}

\newcommand{\xmax}{\ensuremath{X_{\rm max}}}
\newcommand{\nmax}{\ensuremath{N_{\rm max}}}
\newcommand{\ud}{\ensuremath{{\rm d}}}
\newcommand{\nhat}{\ensuremath{\mathbf{\hat{n}}}}
\newcommand{\ehat}{\ensuremath{\mathbf{\hat{e}}}}
\newcommand{\ghat}{\ensuremath{\mathbf{\hat{\gamma}}}}

\makeatletter
\providecommand\add@text{}
\newcommand\tagaddtext[1]{%
  \gdef\add@text{#1\gdef\add@text{}}}%
\renewcommand\tagform@[1]{%
  \maketag@@@{\llap{\add@text\quad}(\ignorespaces#1\unskip\@@italiccorr)}%
}
\makeatother

\begin{document}
\begin{frontmatter}

\title{Universality of Cherenkov Light in EAS}

\author[utah]{Isaac J. Buckland}
  \ead{u0790821@umail.utah.edu}
\author[utah]{D.R.~Bergman\corref{drb}}
  \ead{bergman@physics.utah.edu}

\address[utah]{University of Utah, High Energy Astrophysics Institute,
Salt Lake City, Utah, USA}

\input{abstract}


\end{frontmatter}

\input{introduction}

\input{universality}

\input{cmc}

\input{yield_table}

\input{Cherenkov_Universality_Repo}

\input{EAS_from_table}

\input{CORSIKA_comparisons}

\input{CHASM}

\input{Acknowledgments}

\end{document}

%% file: abstract.tex
\begin{abstract}
    The reconstruction of cosmic-ray-induced extensive air showers with a  non-imaging Cherenkov detector array requires knowledge of the Cherenkov yield of any given air shower for a given set of shower parameters. Although air showers develop in a stochastic cascade, certain characteristics of the particles in the shower have been shown to come from universal probability distributions, a property known as shower universality. Both the energy and the angular distributions of charged particles within a shower have been parameterized. One can use these distributions to calculate the Cherenkov photon yield as an angular distribution from the Cherenkov cones of charged particles at various stages of shower development. This Cherenkov photon yield can then be tabulated for use in the reconstruction of air showers. In this work, we develop the calculation of both the Cherenkov angular distribution and Cherenkov yield per shower particle, and show how a look-up table was constructed to capture the relevant features of these distributions for general use. We compare the results of our calculations with the results of full, particle-stack, Monte Carlo simulation of the Cherenkov light produced in extensive air showers using CORSIKA-IACT. We make comparisons of both the lateral distribution of the Cherenkov photon flux amongst several detectors and of the arrival-time distribution of the Cherenkov photons in a single detector.
\end{abstract}

%% file: introduction.tex
\section{Introduction}
Extensive air showers (EAS) from cosmic rays with energies at or below the knee of the cosmic ray energy spectrum produce a limited amount of fluorescence light. The Cherenkov light produced by such EAS, while similar in total number of photons to fluorescence light, is primarily beamed forward, giving a much larger flux in that direction and permitting the optical detection of EAS at lower energies\cite{TALE}. In a non-imaging Cherenkov detector array, the properties an air shower can be reconstructed using both the time-integrated lateral distribution of the light, as well as the width of the temporal signal, as seen in different detectors of the array. The reconstruction of showers using Cherenkov signals traditionally involves a comparison between data and phenomenologically-determined distributions. A deterministic model of the Cherenkov light distribution from an air shower with given parameters would allow one to perform an Inverse Monte Carlo (IMC) analysis, fitting shower parameters based on collected signals. Such a model will not reproduce the shower-to-shower fluctuations due to hadronic subshowers, and thus may not be useful in the discrimination of photonic from hadronic air showers. The model is conceived in the context of non-imaging Cherenkov detectors at air-shower energies above $10^{14}$ eV.

The plan of this paper is as follows: in Section~\ref{universality} we present a method  to convolve charged-particle energy and angular distributions with the charged-particle Cherenkov to generate a universal, Cherenkov-photon angular distribution. In Section~\ref{cmc} we present a Monte Carlo verification of the analytic solution. In section~\ref{YieldTable} we present a method of tabulating the average Cherenkov photon yield of an air shower. In section~\ref{CherenkovUniversality} we present a public repository where these various calculations have been implemented. In Section~\ref{table} we present a method of using the tabulated Cherenkov angular distribution to reproduce the signal of an air shower in a surface non-imaging Cherenkov array. In Section~\ref{CORSIKA} we present a comparison of the distributions generated to those generated by CORSIKA\cite{CORSIKA} with its IACT extension. Finally, in Section~\ref{CHASM} we describe the CHerenkov Air Shower Model (CHASM), a python module where users can input shower parameters and determine the Cherenkov signal at desired locations.

%% file: universality.tex
\section{Universal Cherenkov Angular Distribution}\label{universality}
Although EAS develop in a stochastic cascade, certain characteristics of the particles in the cascade, including energy and angle, have been shown to represent samples from universal probability distributions \cite{Lafebre,Nerling,Giller}. This is only the case for showers large enough that the distributions of shower-particle properties are meaningful. Both the energy and angular distributions of charged particles are universal in shower stage $t$, and are independent of primary cosmic-ray species \cite{Lafebre}. The charged particles in an EAS will produce Cherenkov light if their energy exceeds the Cherenkov threshold at that point in the atmosphere. The Cherenkov light is produced in a cone of given angle according to the atmospheric index-of-refraction in which they propagate.

The energy of a charged particle in a shower is drawn from a universal distribution $f_e(E_e;t) = \frac{\ud n_e}{\ud l}$ where $t$ is the shower stage and $l=\ln E_e$ with $E_e$ in MeV. For convenience we use $l=\ln E$ to represent the log-energy of secondary particles. The charged particle energy distribution is only dependent on the stage of shower development. To be specific, we will use shower stage as parameterized, $t=\frac{X-X_{\rm max}}{X_o}$\cite{Lafebre}, where $X_o$ is the radiation length of the medium, though other parameterizations could be used such as shower age, $s=\frac{3X}{X+2X_{\rm max}}$\cite{Nerling}.

In the absence of significant geomagnetic field effects, we assume azimuthal symmetry. The angle $\theta_e$ that a charged particle of a given energy makes with the shower axis is drawn from the universal angular distribution $g_e(\theta_e;l_e) = \frac{\ud n_e}{\ud \Omega_e}$. This distribution depends only on the charged particle energy and is independent of stage \cite{Giller}.

\begin{figure}
	\begin{center}
		\includegraphics[width=\columnwidth]{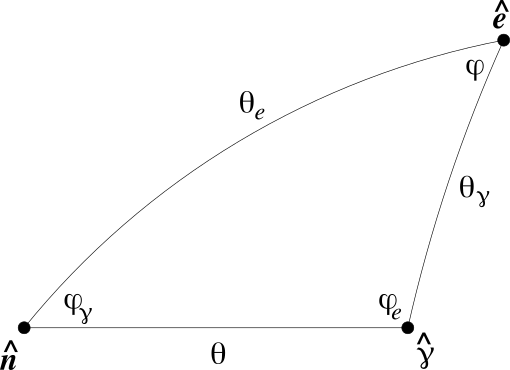}
	\end{center}
	\caption{The angles and directions involved in producing a Cherenkov photon at an angle $\theta$ from the shower axis. The interior ($\phi$) angles are labeled according angle on the opposite side ($\theta$).}
	\label{geometry}
\end{figure}

At a given shower stage, $t$, and atmospheric index-of-refraction, $n=1+\delta$, we wish to calculate the relative Cherenkov photon distribution at an angle $\theta$ from the shower axis direction \nhat. We define this direction to be \ghat. A charged particle of log-energy $l_e$ may emit Cherenkov photons toward \ghat, if its direction from shower axis, \ehat, makes an angle with \ghat\ matching the local Cherenkov cone angle, $\theta_{\rm \check{C}}$. The Cherenkov cone angle is a function of both the electron energy and the local index-of-refraction. The angle between \nhat\ and \ehat\ is denoted $\theta_e$. These angles, and the corresponding interior angles $\phi$, $\phi_e$, and $\phi_\gamma$, form a spherical triangle as shown in Figure~\ref{geometry}.

The number of electrons (charged particles) of a given energy going in direction $\ehat $ is found from the known energy and angular distributions.
\begin{equation}\label{dNe}
    \ud N_e = f_e(E_e; t)\,g_e(\theta_e;E_e)\;\ud l\;\ud\Omega_e
\end{equation}

Charged particles going towards $\ehat $ with $E_e > E_{\rm Th}$, where $E_{\rm Th}$ is the Cherenkov threshold for the given index-of-refraction, will produce Cherenkov photons according to their path length and energy. If we divide out the maximum photon yield of a hyper-relativistic charged particle, $\ud N_\gamma/\ud X (E_e\gg E_{\rm Th})$, we have the relative photon yield 
\begin{equation} \label{yield}
    Y_{\rm \check{C}}(E_e > E_{\rm Th}) = 1-(E_{\rm Th}/E_e)^2
\end{equation}
This is the relative probability that a given charged particle with energy $E_e$ will produce a Cherenkov photon while propagating through a medium with Cherenkov threshold $E_{\rm Th}$. Some of these photons will go into the solid angle about \ghat\ as long as $\theta_\gamma$, the angle a Cherenkov photon could have, matches $\theta_{\rm \check{C}}$. Thus the fraction of photons going towards $\hat{\gamma}$ from electrons going towards \ehat\ is
\begin{equation}\label{dNg}
  \ud N_\gamma = Y_{\rm \check{C}}\,
  \delta(\theta_\gamma-\theta_{\rm \check{C}})\,
  \frac{\ud \Omega_\gamma}{2\pi}
\end{equation}

We convolve the distributions given in equations \ref{dNe} and \ref{dNg} to produce a single relative value of $g_\gamma(\theta;t,\delta) = \frac{\ud n_\gamma}{\ud \Omega_\gamma}$, the angular distribution of Cherenkov photons, at a given $\theta$ away from the shower axis and at a given $t$ and $\delta$. After the integration, the distribution must be normalized because Cherenkov cones from individual electrons simultaneously contribute to multiple increments of $\Omega_\gamma$.
\begin{equation} \label{gg1}
    g_\gamma \propto \int\ud\Omega_e \int\displaylimits_{l_{\rm Th}}^\infty\ud l\ Y_{\rm \check{C}}(l)\,f_e(l)\,g_e(\theta_e;\, l)\,\delta(\theta_\gamma-\theta_{\rm \check{C}})\\
\end{equation}


The key to easily performing the integral is to realize that $\ud\Omega_e$ can be defined in terms of variables related to \ghat\ rather than with respect to \nhat. Thus, $\ud\Omega_e = \sin\theta_\gamma\,\ud\phi_e\,\ud\theta_\gamma$. Also, the domain of $\ud\Omega_e$, which contributes to the solid angle part of the integral above, depends on variables related to \ghat. For a given energy increment, with $\theta$ fixed, the allowed values of $\theta_e$ are found through geometric constraints. When $\nhat \cdot \ehat$ is calculated, the spherical law-of-cosines is recovered for the spherical triangle shown in Figure~\ref{geometry}. 
\begin{equation}\label{loc}
    \cos{\theta_e(\phi_e, \theta_\gamma)} = \nhat \cdot \ehat = \cos{\theta}\cos{\theta_\gamma} + \sin{\theta}\sin{\theta_\gamma}\cos{\phi_e}
\end{equation}
Thus, $g_e$ is a function of $\phi_e$ and $\theta_\gamma$ for a given $l$.

Now we can leverage the delta function to compute the $\theta_\gamma$ part of the angular integral, since for a given $l$, $\theta_{\rm \check{C}}$ is a constant.
\begin{eqnarray}\label{gg_int}
    g_\gamma &\propto&
        \int\displaylimits_{l_{\rm Th}}^\infty\ud l\ Y_{\rm \check{C}}(l)\,f_e(l) \times\\
    && \int\displaylimits_{0}^{2\pi}\ud \phi_e \int\displaylimits_{0}^{\pi}\ud \theta_\gamma\ \sin{\theta_\gamma}\,g_e(\phi_e,\, \theta_\gamma;\, l)\,\delta(\theta_\gamma-\theta_{\rm \check{C}})\\
    &=& \int\displaylimits_{l_{\rm Th}}^\infty\ud l\ \sin{\theta_{\rm \check{C}}(l)}\, Y_{\rm \check{C}}(l)\,f_e(l) \times\\\label{gg_int2}
    && \int\displaylimits_{0}^{2\pi}\ud \phi_e\ g_e(\phi_e,\, \theta_{\rm \check{C}}(l);\, l)
\end{eqnarray}

The double integral in \ref{gg_int2} is performed numerically for a range of shower states and indices-of-refraction and tabulated. This requires specification of the $f_e$ and $g_e$ distributions. We use the distributions given in \cite{Bergman}, but others could be used. The distributions of \cite{Lafebre} should work as well. With others, e.g. \cite{Giller}, care must be taken not to extend the integral into parts of the parameters space where the phenomenologically determined distributions are not valid.

\begin{figure}
	\begin{center}
		\includegraphics[width=\columnwidth]{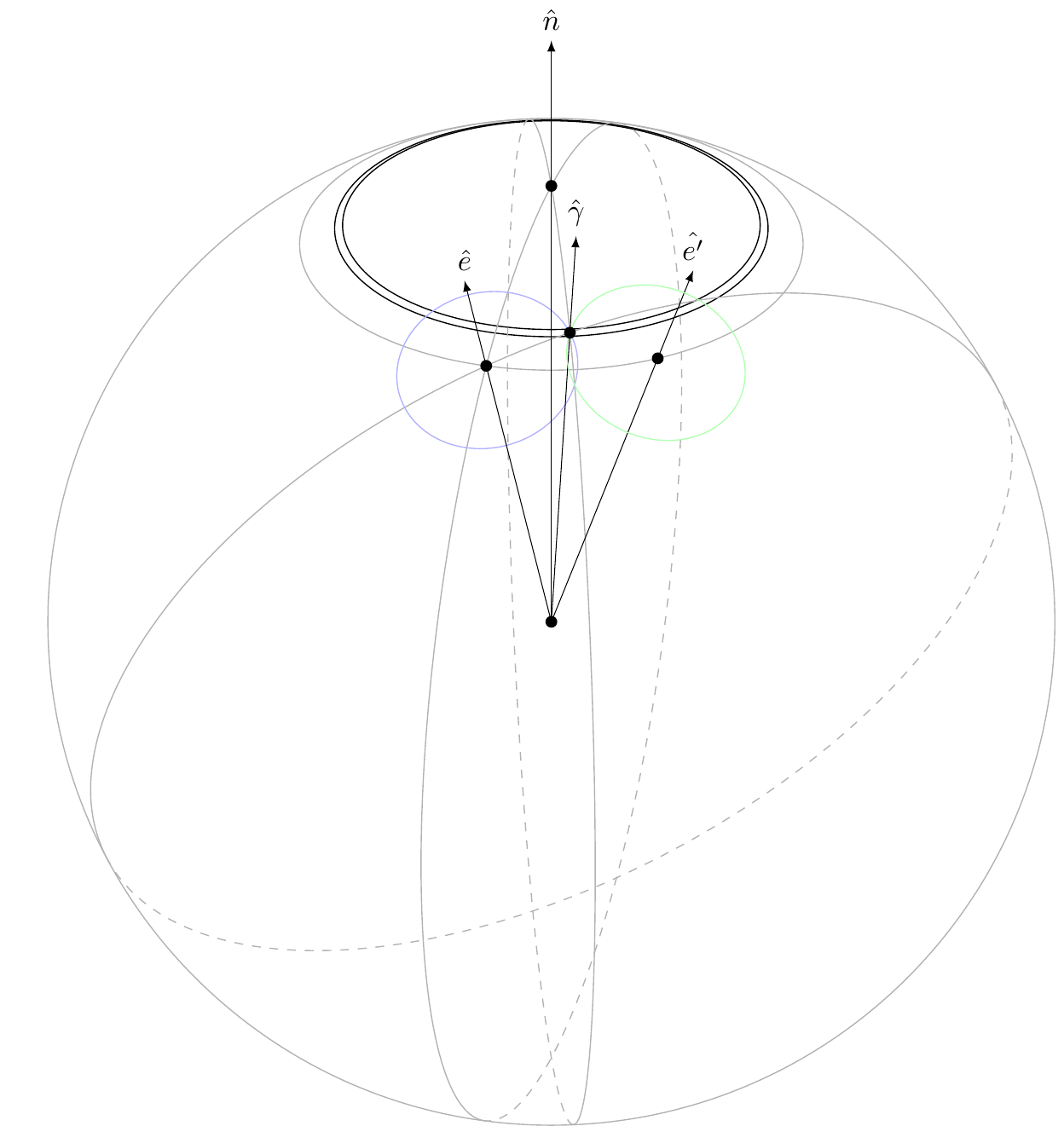}
	\end{center}
	\caption{The projection of the spherical triangle \ref{geometry} onto a unit sphere, showing directions \nhat, \ehat, and \ghat, as well as the Cherenkov cones of two charged particles with the same $\theta_e$.}
	\label{UnitSphere}
\end{figure}

After the integration the distribution is normalized over all solid angle. However, as all charged particle azimuthal angles are equally likely, for every charged particle, there exists a second charged particle with the same $\theta_e$ but a different azimuthal angle $\phi_\gamma$ whose Cherenkov cone can also emit Cherenkov photons towards $\ghat$. This is related to the fact that each charged particle's Cherenkov cone intersects twice with the spherical annulus representing $\ud \Omega$. In other words, equation \ref{loc} has two roots with different values of $\phi_e$. This property implies that the normalization constant is effectively doubled. The projection of the spherical triangle from Figure~\ref{geometry} onto a unit sphere, as well as the Cherenkov cones of equally energetic charged particles, is shown in Figure~\ref{UnitSphere}.

%% file: cmc.tex
\section{Cherenkov Monte Carlo}\label{cmc}

To show the validity of the integration in the previous section, Cherenkov photons were generated via Monte Carlo for comparison. At a given stage of shower development and with a given index-of-refraction (atmospheric $\delta$, which also gives a particular Cherenkov threshold energy), charged particles were drawn from the universal energy distribution $f_e(E_e;t)$ and $g_e(\theta_e;l_e)$. For each of these particles a Cherenkov photon was or was not produced by drawing from the relative Cherenkov yield, equation~\ref{yield}. If produced, a Cherenkov photon was given random azimuthal angle $\phi$. For the generated Cherenkov photons, the angle with respect to the shower axis is now constrained geometrically. In this case, when we take $\nhat \cdot \ghat$ we again recover the spherical law of cosines for figure \ref{geometry}, this time solving for $\theta$.
\begin{equation}\label{mcloc}
    \cos{\theta(\theta_e,\phi, \theta_\gamma)} = \nhat \cdot \ghat = \cos{\theta_e}\cos{\theta_\gamma} + \sin{\theta_e}\sin{\theta_\gamma}\cos{\phi}
\end{equation}

\begin{figure}
	\begin{center}
		\includegraphics[width=\columnwidth]{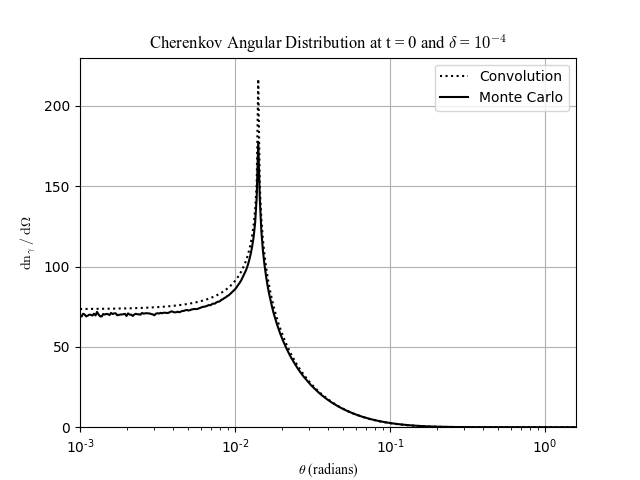}
	\end{center}
	\caption{Comparison between a Cherenkov distribution at $t=0$ and $\delta=10^{-4}$ where the integration is performed by convolution and Monte Carlo respectively. The bins used to collect the Monte Carlo data were chosen to center around the angles tabulated by the convolution integral. While there are slight differences at small angles the peak angle matches exactly.}
	\label{mc_compare}
\end{figure}

The generated $\theta$ values are collected in bins centered around the angles tabulated by the convolution integral, weighted based on the amount of differential solid angle they represent, and the distribution values are normalized. Since the energy distribution only depends on the stage of shower development, the same set of particle energies can be used to compute the Cherenkov distribution of that stage occurring at various heights in the atmosphere. In Figure~\ref{mc_compare} we show the results of throwing $\num{5e8}$ charged particles and compare the distribution to one with the same parameters generated by convolution.

%% file: yield_table.tex
\section{Tabulation of Average Photon Yield}\label{YieldTable}
The number of Cherenkov photons produced by a charged particle in a given propagation interval depends on its energy relative to the local Cherenkov threshold. The photon yield of a charged particle at a given stage $t$ of an EAS and atmospheric $\delta$ is a function of particle energy. Thus, the average value can be found by using the charged particle energy distribution $f_e(E_e, t)$:
\begin{equation}\label{stage_yield}
    \expval{\frac{\ud^2 N_\gamma}{\ud x_e \; \ud N_e}(t,\delta)} = \int\displaylimits_{l_{\rm Th}}^\infty\ud l \; \frac{\ud N_\gamma}{\ud x_e}(E_e,\delta) \; f_e(E_e, t)
\end{equation}
with
\begin{equation} \label{FrankTamm}
    \frac{\ud N_\gamma}{\ud x_e} = 2 \pi \alpha \; \bigg[1 - \frac{1}{(n \beta)^2} \bigg] \; \bigg( \frac{1}{\lambda_{min}} - \frac{1}{\lambda_{max}} \bigg) \;\;\; \left[\gamma \; \mathrm{m^{-1} e^{-1}}\right]
\end{equation}
where $x_e$ is the charged particle propagation interval (in meters), $\alpha$ is the fine structure constant, $\beta$ is the electron's speed in units of $c$ at energy $E_e$, $n$ is the local index of refraction, and $\lambda_{min}$ and $\lambda_{max}$ are the minimum and maximum Cherenkov light wavelengths accepted by the detectors. Since the accepted wavelength interval factor is separable, the value of the rest of the integral has been tabulated at various values of $t$ and $\delta$.

%% file: Cherenkov_Universality_Repo.tex
\section{Cherenkov Universality Repository}\label{CherenkovUniversality}

The numerical methods used to calculate integrals \ref{gg_int2} and \ref{stage_yield} are presented in the Cherenkov Universality python code repository \cite{CherenkovUniversality}. The actual integration is performed in the repository's Cherenkov photon module. It takes on the order of several hours to tabulate Cherenkov distributions for 176 values of $\delta$ and 321 values of $\theta$ at one stage of shower development. Implementations of the charged particle distributions $g_e(\theta_e;l_e)$ and $f_e(E_e;t)$ from \cite{Giller} and \cite{Lafebre} respectively are defined in the repository's charged particle module. Interfaces for these distributions were defined so users could use custom implementations to generate Cherenkov light distributions. These interfaces were made using python's abstract base class support \cite{ABC}.

%% file: EAS_from_table.tex
\section{Calculating EAS Cherenkov Signal from Distribution Tables} \label{table}

For a given EAS trajectory, vectors $\va{r}_a$ from the origin (where the shower axis meets the Earth's surface) to evenly spaced points on the axis are calculated. Based on the locations of supposed photon counters $\va{r}_c$, travel vectors $\va{r}_t = \va{r}_c - \va{r}_a$ from the axis to the counters are calculated. A shower profile, either given directly or calculated from parameters, as a function of slant depth is then assigned to the axis based on the depth intervals between axis points. A shower development stage $t$ is assigned to each axis point according to this profile. An atmospheric delta $\delta$ is also calculated based on the altitude of each sampled axis point.

The total number of Cherenkov photons produced per meter per charged particle at a shower stage is found by accessing the pre-compiled table described above in Section~\ref{YieldTable}. To first order, we assume that the distance each charged particle travels during each sampled stage of shower development is similar to the spatial interval represented by each corresponding axis interval $\ud r_a \approx \ud x_e$. We multiply the spatial intervals $\ud r_a$ by the tabulated yields per particle per meter from equation \ref{stage_yield}. This is the total number of Cherenkov photons produced over all solid angles at each sampled stage. 

The fractions of photons going from the specific point in the EAS toward specified counting locations are found by sampling the table of angular distributions described in Section~\ref{universality}. Based on the shower stages $t$, as well as the atmospheric delta $\delta$ at each axis point, a Cherenkov angular distribution is calculated via interpolation from the table. This distribution is sampled at the angles which the vectors $\va{r}_t$ make with the shower axis.

Before factoring in detector response and acceptance, each counter simply represents an amount of solid angle $\ud \Omega$ as seen by the production point on the axis. When multiplied by the photon density found from the tables, we find the total number of photons arriving from a specific axis point. 
\begin{equation}\label{Ng}
    N_\gamma(\theta, \; t, \; \delta) = 2 \; \expval{\frac{\ud^2 N_\gamma}{\ud x_e \; \ud N_e}(t,\delta)} \; N_e(t) \; g_\gamma(\theta; \; t, \; \delta) \; \ud \Omega \; \ud r_a
\end{equation}


%% file: CORSIKA_comparisons.tex
\section{Comparing to CORSIKA IACT}\label{CORSIKA}
\begin{figure}
	\begin{center}
		\includegraphics[width=\columnwidth]{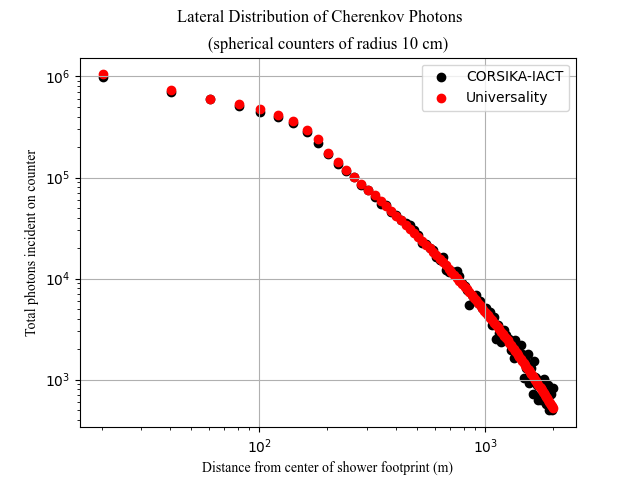}
	\end{center}
	\caption{Comparison of Cherenkov lateral distribution, the total number of photons collected from the whole shower at each spherical counting volume, from both CORSIKA and universality.}
	\label{LateralDistribution}
\end{figure}

\begin{figure}
	\begin{center}
		\includegraphics[width=\columnwidth]{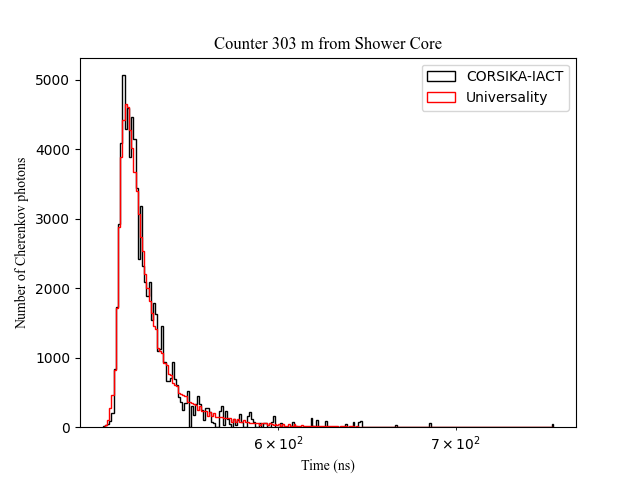}
	\end{center}
	\caption{Comparison of arrival time distribution of Cherenkov photons at just one counter from both CORSIKA and universality. The fluctuation in the CORSIKA IACT signal is due to the concentration of charged particles into sub-showers.}
	\label{ArrivalTimes}
\end{figure}

To demonstrate the veracity of this Cherenkov universality model, showers were generated using CORSIKA's IACT extension, with IACT's defined at increasing distances from the shower core \cite{CORSIKA}. The particular shower shown in figures \ref{LateralDistribution} and \ref{ArrivalTimes} is a proton shower with a primary energy of $10^8$ GeV, \xmax\ of $666 \, \textrm{g/cm}^2$, \nmax\ of $\num{6.3e7}$ particles, and a zenith angle of 30$^\circ$. \xmax\  and \nmax\  refer to the atmospheric depth at shower maximum and the maximum number of secondary shower particles, respectively. The observation level was defined at sea level. Shower development stages were calculated at each step in slant depth based on CORSIKA's longitudinal profile compared to \nmax. Atmospheric $\delta$s along the profile were calculated via interpolation from depth along the axis as a function of altitude. The U.S.\ Standard Atmosphere of 1976 was used for all altitude and density calculations.

Using the method described in section \ref{table}, Cherenkov signals were calculated along this shower's longitudinal profile toward each counting location. We compare this signal to that generated by CORSIKA IACT. The total number of photons collected at increasing distances from the shower core are compared in Figure~\ref{LateralDistribution}.

When considering Cherenkov production, CORSIKA's clock starts when the primary particle enters the atmosphere \cite{CORSIKA}. To calculate when our Cherenkov signals arrive at each counting location, we approximate the shower front as traveling along the axis at $c$, thus calculating the time it takes for the shower front to move from the top of the atmosphere to each axis location $\va{r}_a$. We then compute the amount of time it would take for something moving at $c$ to travel along travel vectors $\va{r}_t$. This time is adjusted by the delay the Cherenkov photons experience as they propagate along $\va{r}_t$. A vertical delay is calculated as the sum of delays through refractive indices at intervals surrounding each sampled height, then divided by the cosine of the polar angle of vector $\va{r}_t$. The arrival time distributions for a counter 303 m from the shower core are compared in Figure~\ref{ArrivalTimes}.

%% file: CHASM.tex
\section{CHASM (CHerenkov Air Shower Model)}\label{CHASM}
CHASM is a python module which allows users to input shower parameters and counter locations, and a Cherenkov signal is generated using the methods described in this article. It uses methods which account for the curvature of the atmosphere for showers with large zenith angles. By default, the U.S.\ Standard Atmosphere of 1976 is used, however any atmospheric model can be used using the atmosphere interface. Tables generated by the Cherenkov Universality repository \ref{CherenkovUniversality} are included in the CHASM repository as a compressed NumPy file. Users can also use their own table by simply replacing this file as long as the included arrays have the same names and ranks. Compared to CORSIKA IACT, which may take up to many hours to generate a single high-energy air shower from which Cherenkov signals are extracted, CHASM can produce the signals from the same shower in seconds.

The goal of CHASM is to apply Cherenkov universality to a wide range of air shower observation scenarios. The first iteration of CHASM is being tested for implementation in nuSpaceSim, a comprehensive neutrino simulation package for space-based and suborbital experiments \cite{NuSpaceSim}. Tau neutrinos skimming the Earth may interact via the charged-current interaction in the Earth’s crust. The resulting tau particle leaves the Earth and decays, serving as the primary particle in an upward going air shower. As direct simulation of upward going Cherenkov light becomes available in CORSIKA 8 \cite{CORSIKA8}, further development and comparisons will be possible.

The module is available for use at \cite{CHASM}. There are examples which describe how to use it to generate EAS signals, and even EAS libraries. We include an example comparing the signal generated by CHASM directly to that of CORSIKA IACT. It also contains the methods needed to generate unique yield and angular distribution tables with different parameterizations for both the particle energy and angular distributions.

%% file: Acknowledgments.tex
\section{Acknowledgments}\label{Acknowledgments}
This work was supported by NSF grants PHY-1806797, PHY-2112904, and PHY-2209583, as well as NASA grant 80NSSC19K0485 at the University of Utah. I would like to thank Dr. Yoshiki Tsunesada at Osaka Metropolitan University for his helpful suggestions. An allocation of computer time from the Center for High Performance Computing at the University of Utah is gratefully acknowledged.